# Quasilocalized Vibrational Modes as Efficient Heat Carriers in Glasses

Xing Xiang[1], Sylvain Patinet[2], Sebastian Volz[3] and Yanguang Zhou[1*]

[1]*Department of Mechanical and Aerospace Engineering, The Hong Kong University of Science and Technology, Clear Water Bay, Kowloon, Hong Kong SAR, China*

[2]*PMMH, CNRS, ESPCI Paris, Université PSL, Sorbonne Université, Université de Paris, 7 quai St Bernard, Paris, 75005, France*

[3]*LIMMS/CNRS-IIS(UMI2820) Institute of Industrial Science, University of Tokyo 4-6-1 Komaba, Meguro-ku Tokyo, 153-8505, Japan*

## Abstract

While soft quasilocalized vibrational modes are known to populate the low-frequency spectrum of glassy solids, their contribution to thermal properties is still not fully elucidated. We numerically show that, despite their spatially localized nature, these modes are as effective heat carriers as the delocalized ones and can contribute non-negligible to the total thermal conductivity in the low-temperature regime, especially for $T < 0.05\ T_g$ where $T_g$ is the glass transition temperature. We further prove that the mutual coherence between the low-frequency quasilocalized modes and other modes explains this high thermal exchange performance. Our finding finally provides a perspective on the thermal transport behaviour of the low-frequency quasilocalized modes in glassy solids.

---

[*]Author to whom all correspondence should be addressed. E-Mail: maeygzhou@ust.hk (Y.Z.)

The low-temperature thermodynamic properties of dielectric solids, e.g., heat capacity and thermal conductivity, depend on the atomic vibrations at low frequencies. For example, in crystals, the specific heat increases as the third power of temperature stemming from the vibrational density of states (*VDOS*) being proportional to the angular frequency ($\omega$) squared [1,2]. The same principle also holds in glassy solids, where the low-frequency vibrations determine the low-temperature thermodynamic properties [3,4], however, according to a significantly different mechanism from that of the crystal. Theoretical considerations have pointed out that anomalous low-frequency vibrational modes in three-dimensional (3D) glasses are at the origin of the broad boson peak in the reduced vibrational density of states $VDOS(\omega)/\omega^2$ [5-7], leading to an anomalous temperature dependence of the heat capacity and thermal conductivity [3,4,8,9]. It has been highlighted that a fraction of the vibrational modes residing in the low-frequency region are quasilocalized and originate from the coupling between disordered structures and long-wavelength transverse excitations [10]. These modes present a density of states with a universal dependence on frequency, $\omega^4$ [11-18]. Unlike delocalized vibrational modes, which are expected to be efficient heat carriers in solids [19-22], the energy of localized modes has been shown to be spatially remain where they reside [22-27]. These low-frequency quasilocalized vibrational modes (QVMs) are then presumed to be inefficient heat carriers and their contribution to thermal conductivity is considered as negligible. It has recently been shown that the contribution of some modes like QVMs to thermal exchange is non-negligible in glassy polymer systems [28]. However, the underlying mechanism behind this unexpected heat transfer is unclear. A first principle understanding of the thermal transport behaviour of these QVMs is of key importance in determining the low-temperature thermodynamic properties of glasses. Using binary glassy models and quasi-harmonic Green-Kubo (QHGK) calculations, we directly prove that QVMs are actually very efficient heat carriers and contribute to most of the thermal conductivity at low temperatures, especially when $T < 0.05\ T_g$ where $T_g$ is the glass transition temperature. The high heat transport capacity of these modes is found to originate from the strong coherence effects between them and other vibrational modes.

We employ the binary system described in [29] where the two types of atom interact via a modified Lennard-Jones (LJ) interatomic potential with reduced units of mass $m$, energy $\varepsilon$ and length $\sigma$. Accordingly, the unit of time is defined as $t_0 = \sigma\sqrt{m/\varepsilon}$ and frequency $f$ is measured in the unit of $t_0^{-1}$. With the Boltzmann constant $k_B$, the units of temperature, heat capacity and thermal conductivity can be expressed as $\varepsilon/k_B$, $k_B$ and $k_B/t_0\sigma$, respectively. The number ratio between the two types of atoms is $(1+\sqrt{5})/4$. The simulations are carried out at constant volume with a density equal to unity and periodic boundary conditions in all three directions. The computer glass transition temperature $T_g$ is located around $0.184\ \varepsilon/k_b$. The numerical glass samples are generated with classical molecular dynamics (MD) from liquids equilibrated at $1.908\ T_g$ and then instantaneously quenched. After a potential energy minimization, the amorphous structures are then relaxed for $18067\ t_0$ in the isothermal-isobaric ensemble at $0.163\ T_g$ and zero pressure, and once again relaxed at equilibrium. The detailed procedure to generate the binary glassy model can be found in Refs. [29-31]. A normal mode analysis is performed from the calculation of eigenfrequencies $\omega$ and eigenvectors $\vec{e}$, which satisfies $\sum_j \mathbf{D}_{ij}\vec{e}_j = \omega^2 \vec{e}_i$, where $\mathbf{D}_{ij}$ is the dynamical matrix. The *VDOS* and participation ratio (*PR*) are calculated with $VDOS(\omega) = \sum_{\lambda=1}^{3N-3} \delta(\omega - \omega_\lambda)/(3N-3)$ and $PR = \left[ N \sum_i (\vec{e}_i \cdot \vec{e}_i)^2 \right]^{-1}$ [32], respectively, where $\omega_\lambda$ is the angular frequency of the vibrational mode $\lambda$ and $N$ is the number of atoms. The thermal conductivity $K$ of the glassy systems is deduced from the quasi-harmonic Green-Kubo (QHGK) method [33], which considers both diagonal and off-diagonal elements of the heat current operators [23,34] according to the following expression $\kappa = \frac{1}{V} \sum_{\lambda,\lambda'} C(\lambda, \lambda') D(\lambda, \lambda')$ [33,35], where $V$ refers to the volume of the system, $D(\lambda, \lambda')$ and $C(\lambda, \lambda')$ denote the generalized diffusivity and heat capacity of the $(\lambda, \lambda')$ coherent modes, respectively. Details about the QHGK method can be found in Supplemental Material (SM) [36].

We distinguish the low-frequency QVMs from the delocalized modes based on their *PR*s which are widely used to evaluate modal localization degrees [12-14,32,44]. The *PR* is in the order of $O(1)$ for fully delocalized vibrational modes and of $1/N$ for vibrations localized on an atom. In the following, we select a limit for the participation ratio ($PR = 0.2$) as a criterion to distinguish the delocalized vibrational modes from localized ones or QVMs (i.e., $PR < 0.2$). We note that the cut-off $PR = 0.2$ is chosen as an *ad hoc* value [14,45-47]. **Figure 1a** shows the *PR* as the function of the mode frequency for the different system sizes. It can be found that the upper-frequency limit for the region fully dominated by QVMs approaches ~0.166 $\iota_0^{-1}$ for systems with 512 atoms, whereas this limit reaches ~0.173 $\iota_0^{-1}$ for the systems with 1000 atoms and 2197 atoms, respectively. This threshold presumably underestimates the number of QVMs but provides a safe estimate as QVMs with larger frequencies may hybridize with similar frequency delocalized modes [14,17,32,36,48-50]. The $VDOS(\omega)$ reported in **Figure 1b** appears to follow the $\omega^4$ scaling law as expected for the QVMs in glassy systems [11-18]. As frequency increases, emerging delocalized vibrational modes tend to appear, and the proportion of the QVMs decreases gradually, which makes the $VDOS(\omega)$ deviate from the $\omega^4$ dependence. We also investigated the effect of the quench rate of the preparation protocol on the QVMs [36]. In agreement with [13,17], we observe that the VDOS scaling and the presence of low-frequency vibrational modes with small $PR$ are unaffected by the preparation protocol. In **Figure 1c**, we plot the eigenvector of a quasilocalized mode with $PR \approx 0.037$ as an example, corresponding to the smallest frequency ($0.059 \ t_0^{-1}$) from a typical configuration with 512 atoms. The $PR$ value of 0.037 implies that, on average, only 19 among 512 atoms are involved in this mode. The visualization of the eigenvector highlights that the thermal displacement of a few atoms is much larger than that of the others, and that the corresponding mode is localized [13,28,51].

We then turn to the thermal transport properties of the glassy systems. The total thermal conductivity calculated using the QHGK method is firstly compared with the one computed from the Green-Kubo equilibrium MD (GKEMD) simulations for the validation purpose [36]. The QHGK with the Bose-Einstein distribution is thereafter used in the following calculations. **Figure 2a** shows the modal thermal conductivity of the $N$=512 glasses, which is calculated by $\kappa(\lambda) = \frac{1}{V} \sum_{\lambda'} C(\lambda, \lambda') D(\lambda, \lambda')$. The generalized heat capacity $C(\lambda, \lambda')$ caused by the interaction between two modes $\lambda$ and $\lambda'$ can be regarded as the energy gained among these two modes when the temperature increases since the generalized heat capacity proposed in [36] takes the form $C(\lambda) + \omega_\lambda \left[ C(\lambda) - C(\lambda') \right] / (\omega_{\lambda'} - \omega_\lambda)$. Our results show that the contribution of QVMs and most of the delocalized modes appear as much larger than that of the high-frequency localized ones. We emphasize that all the vibrational modes include the single mode contribution (diagonal term, i.e., $\kappa(\lambda) = \frac{1}{V} C(\lambda, \lambda) D(\lambda, \lambda)$) and the coherence contribution (off-diagonal term, i.e., $\kappa(\lambda) = \frac{1}{V} \sum_{\lambda' \neq \lambda} C(\lambda, \lambda') D(\lambda, \lambda')$) [52]. We distinguish the single mode contribution using $\left| f(\lambda) - f(\lambda') \right| \leq \Delta f$, in which $\Delta f$ is the smearing frequency with a value of 0.042 $t_0^{-1}$ and is equal to the smearing energy in the modal linewidth calculation [36]. **Figure 2b** shows the single-mode and coherence contribution of all the vibrational modes. While some modes with frequency larger than the QVM upper limit have *PRs* smaller than 0.2, their single-mode contribution is comparable to the coherence contribution (**Figure 2b**), which may be caused by their hybridization with similar frequency delocalized modes [14,17,32,36,48-50]. We therefore regard these modes as hybridization modes. The high-frequency localized modes and low-frequency QVMs contribute to the thermal energy exchange mainly through the coherence between them and other modes. In the case of delocalized modes and hybridized modes, we find that both diagonal and off-diagonal terms contribute to the thermal energy exchange. We further estimate the average thermal conductivity of the low-frequency QVMs and compared it with the thermal conductivities of other modes. It is remarkable that the average thermal conductivity of $4.46 \square 10^{-3}$ $k_\mathrm{B}/t_0\square$ accounted to the low-frequency QVMs is close to the modal thermal conductivity of most low-frequency delocalized vibrational modes below $1\,t_0^{-1}$ (**Figure 2b**). This outcome seems to contradict the common view that the thermal energy of the quasilocalized or localized vibrational modes does not propagate [22-27].

We now move to unveil the underlying mechanism that causes the unexpected efficient heat transfer of these QVMs. As mentioned above, the modal thermal conductivity contributed by the off-diagonal terms quantifies the mutual coherence between a specific mode with all other vibrational excitations. To compare the thermal properties of the low-frequency quasilocalized, delocalized and high-frequency localized modes, we select three typical excitations with frequencies of $f_{\lambda_1}=0.121\,t_0^{-1}$, $f_{\lambda_2}=0.552\,t_0^{-1}$, and $f_{\lambda_3}=4.464\,t_0^{-1}$. Among these, the low-frequency quasilocalized and high-frequency localized vibrational modes have a similar *PR* value of ~0.062, while the *PR* value of a delocalized mode is 0.322. Our results show that the heat capacity gain between frequencies $f_{\lambda_1}$ ($f_{\lambda_2}$) and other modes is as substantial as 0.9 $k_B$ (0.6 $k_B$) (**Figures 3a** and **3c**), where $k_B$ is the theoretical maximum for a single mode. However, the heat capacity gained via coherence between $f_{\lambda_3}$ and other modes decreases quickly to zero as the frequency of that mode increases (**Figures 3a** and **3c**). This effect arises from the low occupation number of that mode at 0.163 $T_g$. We further investigate the generalized diffusivity $D(\lambda,\lambda')$ spectrum, which can be interpreted as the thermal diffusivity resulting from the coherence between modes $\lambda$ and $\lambda'$. The generalized diffusivity spectrum uncovers that the coherence between a mode and its neighbouring ones is significant (see the red region in **Figure 3b**) and decreases when the difference between their respective frequency becomes larger. **Figure 3d** quantitatively proves that the generalized diffusivity for two modes with a small difference in frequency is significant, which indicates a strong coherence between those two modes. We further observe that the coherence for all the possible modes contributed by the off-diagonal terms, including low-frequency quasilocalized, delocalized and high-frequency localized modes with close frequencies beyond the smearing energy is important, which leads to the large generalized diffusivity. The low thermal conductivity of these high-frequency modes shown in **Figure 2a** is caused by the lower generalized heat capacity and diffusivity than those of high modes.

We finally focus on the contribution of the low-frequency QVMs to the total thermal conductivity. While the modal thermal conductivity of the low-frequency QVMs is comparable to that of the delocalized vibrational modes (**Figure 2b**), their contribution to the total thermal conductivity remains negligible when most of the modes are excited (**Figure 4a** and **4b**). This outcome lies in the low number ratio of the low-frequency QVMs (**Figures 1a** and **1b**). The low-temperature thermodynamic properties are determined by the low-frequency modes as only these modes are occupied. As a result, the contribution to total thermal conductivity from the low-frequency QVMs varies from ~2% to ~30% in the temperature range below 0.05 $T_g$ (**Figures 4a** and **4b**). When the temperature of the system increases, more delocalized modes with higher frequencies are excited and contribute to thermal transport. The contribution to the total thermal conductivity from the middle-frequency delocalized modes and the high-frequency ones increases with temperature and becomes dominant (**Figures 4a** and **4b**). **Figures 4c** and **4d** show the temperature dependence of the specific heat capacity and the thermal conductivity contributed by QVMs, respectively. Below ~0.01 $T_{\mathrm{g}}$, the specific heat capacity presents a clear ~$T^5$ relation, which is stemming from the ~$\omega^4$ scaling law of VDOS and in accordance with the theoretical predictions [1,9]. The corresponding thermal conductivity at low temperatures then shows an unusual ~$T^6$ relation (**Figure 4d**). It is noted that the thermal conductivity of the real glassy systems at low temperatures shows a temperature dependence of ~$T^2$ scaling [3,4,8], which is different from our results here. In real glassy systems, low-frequency QVMs, the low-frequency single vibrational modes [1] and two-level system [11,53,54] contribute to the thermal energy transport. The low-temperature dependence of the thermal conductivity in some real glassy systems is then determined by the dominant heat carriers. For instance, the thermal conductivity of amorphous silica at low temperature follows the ~$T^2$ scaling relation since the single vibrational modes are the main heat carriers at the sub-Terahertz frequency region [8,55]. While the glassy systems in our simulations are designed to only include the QVMs at the extremely low-frequency region, it is possible to introduce these ultralow frequency single modes via increasing the size of the system [12]. Consequently, the temperature dependence of the thermal conductivity at low temperatures will derivate from the ~$T^2$ relations as observed in some real glassy systems [8]. This may give a guide to experimentally observe the QVMs.

In conclusion, this letter highlights that the QVMs in bulk glassy solids are as efficient heat carriers as delocalized vibrational modes and can contribute non-negligible to the total thermal conductivity in the low-temperature range. The spectral analysis further proves that the mutual coherence and the heat capacity increase due to the coupling between the low-frequency QVMs and other modes are the reasons for the high thermal exchange performance of these low-frequency QVMs. We finally uncovered an unexpected coherence mechanism, which has a broad impact on the thermal behaviour of glassy solids.

**Acknowledgements**

Y.Z. thanks the start-up fund (a/c-R9246), SJTU-HKUST joint research collaboration fund (SJTU21EG09), the Bridge Gap Fund (BGF.008.2021) from the Hong Kong University of Science and Technology (HKUST) and the Hong Kong SciTech Pioneers Award from the Y-LOT Foundation.

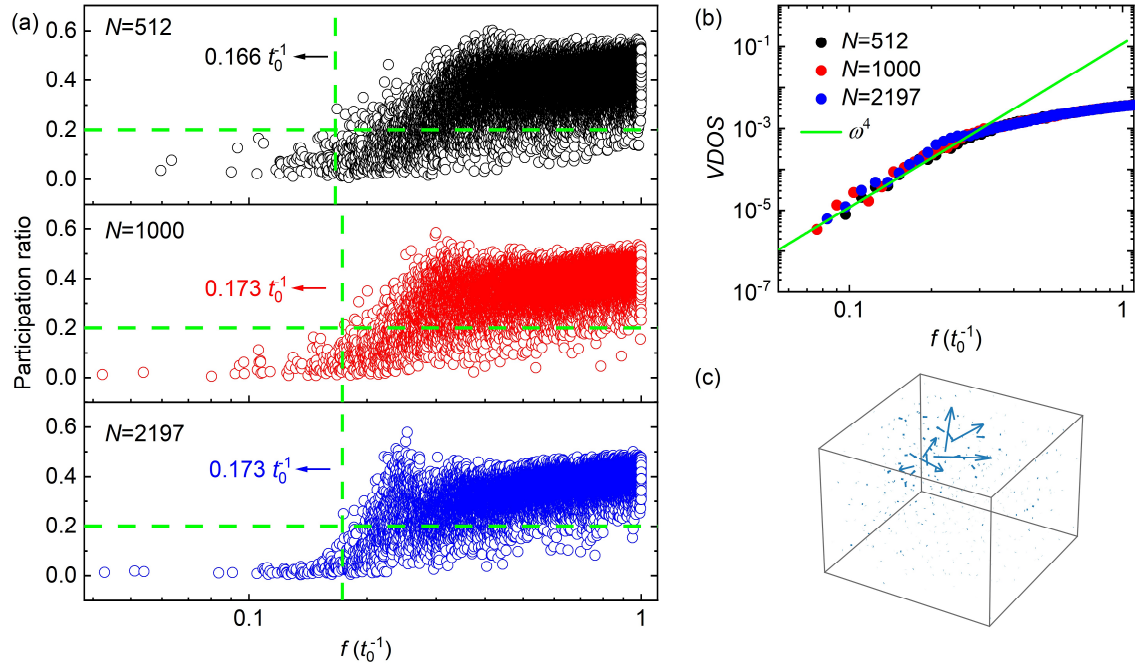

**Figure 1**. Characteristics of QVMs. (a) Modal *PR* below 1 $t_0^{-1}$ as the function of frequency. (b) Vibrational density of states (*VDOS*) of the three glassy systems of different sizes. (c) Visualization of the eigenvectors for the mode with $f = 0.059$ $t_0^{-1}$ and $PR = 0.037$ for a 512-atom system. The arrow length in (c) is set according to the modulus of the normalized eigenvectors and magnified by a factor of ten for visualization. The results in (a) and (b) are obtained based on 250, 100 and 50 samples for 512-atom, 1000-atom and 2197-atom systems, respectively.

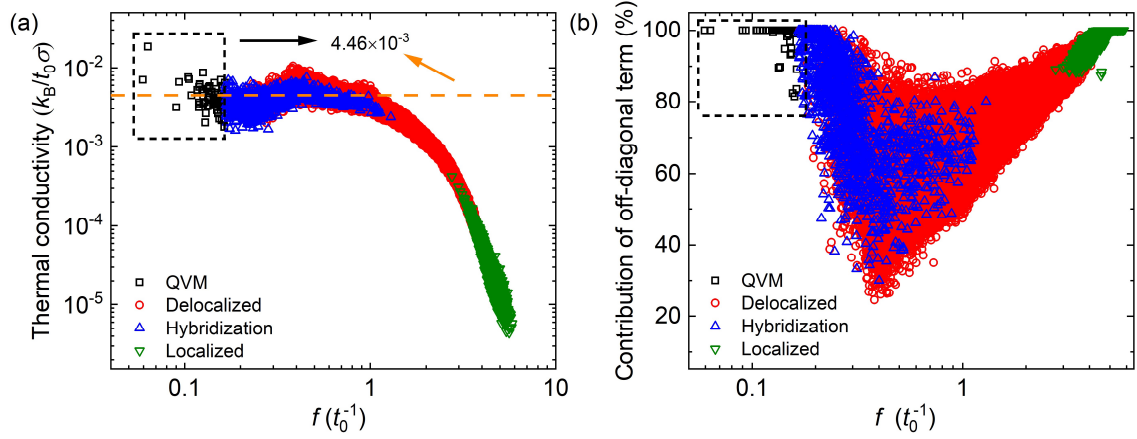

**Figure 2**. Modal thermal conductivity of glassy systems for QVM (black), delocalized (red),

hybridization (blue), and localized (green) modes. (a) The modal thermal conductivity for systems with 512 atoms, is calculated by the QHGK method which assumes that vibrations follow the Bose-Einstein distribution. The horizontal dashed orange line corresponds to the averaged modal thermal conductivity $4.46\times10^{-3}$ $k_B/t_0\sigma$ of QVMs. All the data are obtained at the temperature of 0.163 $T_g$. (b) The contribution of off-diagonal elements to the thermal transport for every mode in (a).

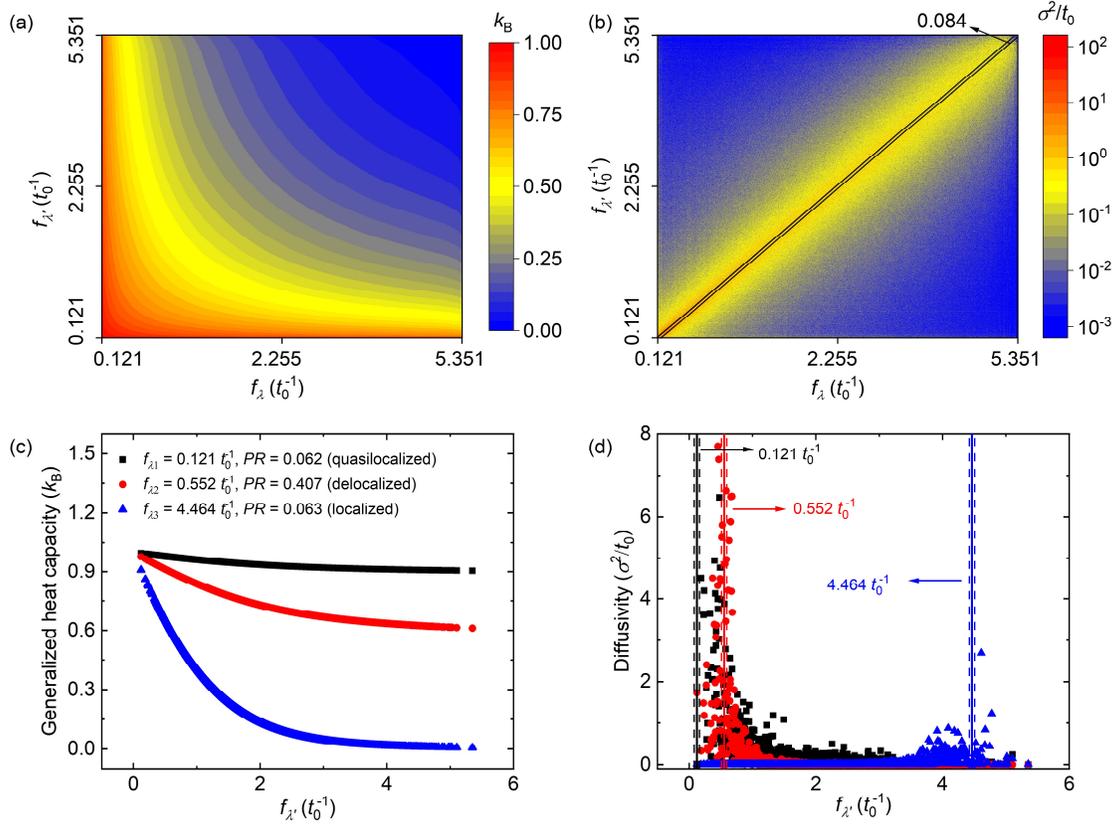

**Figure 3**. Generalized heat capacity (left column) and generalized diffusivity (right column) using a 512-atom configuration at 0.163 $T_g$ for all modes in (a) and (b) and for three typical vibrational modes in (c) and (d). Three modes are $f_{\lambda_1} = 0.121 t_0^{-1}$ with $PR = 0.062$ (quasilocalized), $f_{\lambda_2} = 0.552 t_0^{-1}$ with $PR = 0.407$ (delocalized) and $f_{\lambda_3} = 4.464 t_0^{-1}$ with $PR = 0.063$ (localized). The distance between two solid lines is twice of the smearing energy (0.084 $t_0^{-1}$), which is the width of the energy conservation smearing in the phonons scattering calculation.

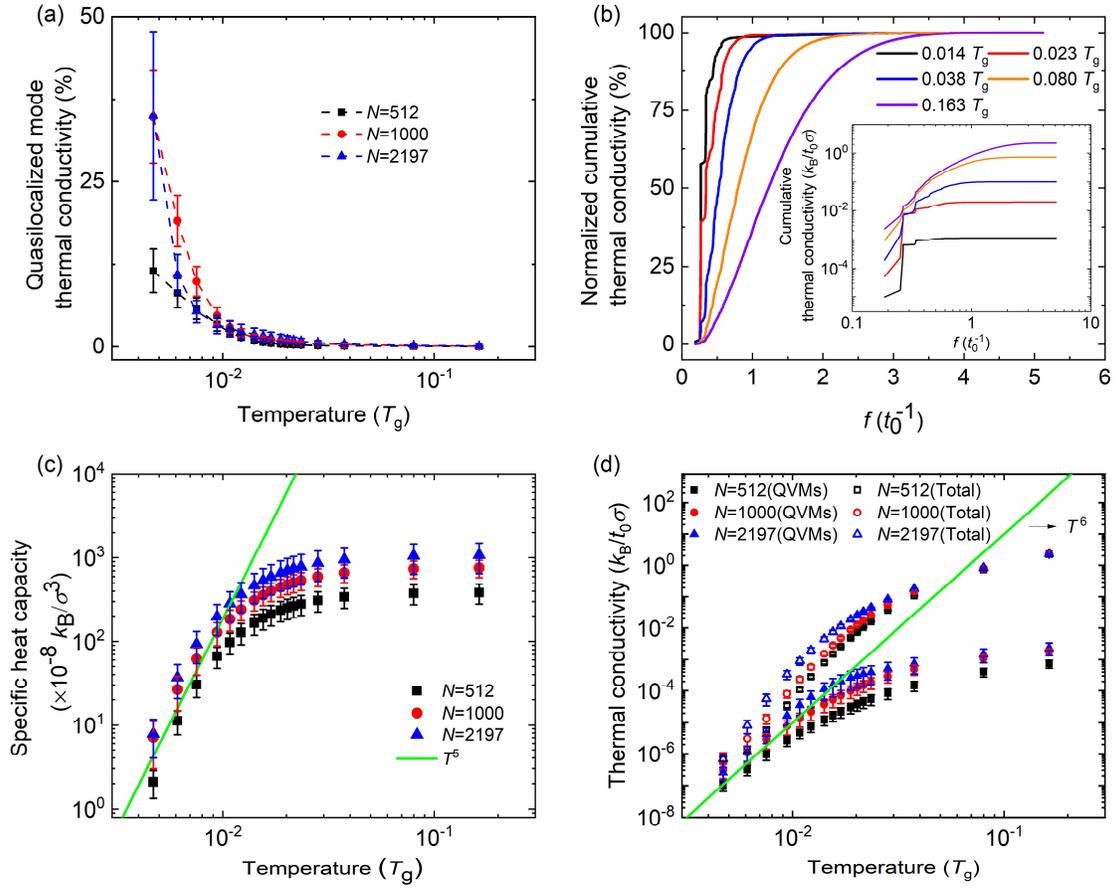

**Figure 4.** (a) Contribution to the total thermal conductivity of these low-frequency QVMs for three glassy systems with 512, 1000 and 2197 atoms. (b) The normalized cumulative thermal conductivity function vs. frequency for a 512-atom configuration at five typical temperatures. The corresponding cumulative thermal conductivity is also shown in the inset. (c) The specific heat capacity as the function of temperature. (d) The total and QVMs contributed thermal conductivity versus temperature. The results in (a), (c) and (d) are averaged over 60, 20 and 5 configurations for 512-atom, 1000-atom and 2197-atom systems, respectively.